\documentclass[showpacs,preprintnumbers,amsmath,amssymb,superscriptaddress,prd,nofootinbib]{revtex4}
\usepackage{graphics}
\usepackage{epsfig}                                                            
\usepackage{time}
\usepackage{bm}
\newcommand{\NIM}[3]       {Nucl.\ Instr.\ Methods~{\bf A#1} (#2) #3}

\newcommand{\NPB}[3]       {Nucl.\ Phys.~{\bf B#1} (#2) #3}
\newcommand{\PLB}[3]       {Phys.\ Lett.~{\bf B#1} (#2) #3}

\newcommand{\PRD}[3]       {Phys.\ Rev.~{\bf D#1} (#2) #3}
\newcommand{\PRL}[3]       {Phys.\ Rev.\ Lett.~{\bf #1} (#2) #3}

\newcommand{\EPJ}[3]       {Eur.\ Phys.\ J.~{\bf C#1} (#2) #3}
\newcommand{\CPC}[3]       {Comp.\ Phys.\ Comm.~{\bf #1} (#2) #3}
\newcommand{\journal}[4]   {#1~{\bf #2} (#3) #4}
\newcommand{\hb} {\mbox{HERA\protect\rule[.5ex]{1.ex}{.11ex}B}\ }
\newcommand{\hbp}{\mbox{HERA\protect\rule[.5ex]{1.ex}{.11ex}B}}
\newcommand{\ra} {\mbox{$\mskip 3mu \rightarrow \mskip 5mu$}}

\newcommand{\etal} {{\it et~al.}}

\newcommand{\eff}{\ensuremath{\varepsilon}}
\newcommand{\effB}{\ensuremath{\eff_{\bbar}^{\jpsi}}}
\newcommand{\effBz}{\ensuremath{\eff_{\bbar}^{\Delta z}}}

\newcommand{\effP}{\ensuremath{\eff_P^{\jpsi}}}

\newcommand{\effR}{\ensuremath{\eff_R}}
\newcommand{\bbjX}{\ensuremath{\bbar\ra\jpsi X}}

\newcommand{\Br}[1]{\ensuremath{{\rm  Br}(#1)}}
\newcommand{\sigB}{\ensuremath{\sigma_{\bbar}^A}}
\newcommand{\sigP}{\ensuremath{\sigma_P^A}}
\newcommand{\dsigB}{\ensuremath{\Delta\sigma_{\bbar}^A}}
\newcommand{\dsigP}{\ensuremath{\Delta\sigma_P^A}}
\newcommand{\nP}{\ensuremath{n_{P}}}
\newcommand{\nB}{\ensuremath{n_{\bbar}}}
\newcommand{\fP}{\ensuremath{f_{P}}}
\newcommand{\fB}{\ensuremath{f_{\bbar}}}

\newcommand{\Dz}{\ensuremath{\Delta z}}

\newcommand{\PBp}  {\mbox{\ensuremath{{B}^+}}}                   
\newcommand{\PBz}  {\mbox{\ensuremath{{B}^0}}}                   

\newcommand{\PKp} {\mbox{\ensuremath{{K}^+}}}                    

\newcommand{\Bee}{\ensuremath{{b}}}
\newcommand{\bjpsiX}{\ensuremath{ \Bee\to\jpsi X}}
\newcommand{\bjpsi}{\ensuremath{\Bee \to\jpsi}}

\newcommand{\bjkp}{\ensuremath{\PBz \to\jpsi \PKp \pi^-}} 
\newcommand{\bjk }{\ensuremath{\PBp \to\jpsi \PKp}}       

\newcommand{\sigbbar}{\ensuremath{\sigma(\bbar)}}
\newcommand{\dsigbbar}{\ensuremath{\Delta\sigma(\bbar)}}
\newcommand{\bbar}{\ensuremath{\Bee\overline{\Bee}}}
\newcommand{\ccbar}{\ensuremath{c \overline{c}}}

\newcommand{\ee}{\ensuremath{e^+e^-}}
\newcommand{\mm}{\ensuremath{\mu^+\mu^-}}

\newcommand{\jpsi}{\ensuremath{J/\psi}}

\newcommand{\jpsill}{\ensuremath{\jpsi\to l^+l^-}}
\newcommand{\jpsiee}{\ensuremath{\jpsi\to e^+e^-}}
\newcommand{\jpsimm}{\ensuremath{\jpsi\to \mu^+\mu^-}}
\newcommand{\xf}{\ensuremath{x_{ F}}}
\newcommand{\pt}{\ensuremath{p_{ T}}}

\newcommand{\egev}{\ensuremath{\, \mathrm{GeV}}}
\newcommand{\mgev}{\ensuremath{\, \mathrm{GeV}/c^2}}
\newcommand{\pgev}{\ensuremath{\, \mathrm{GeV}/c}}

\newcommand{\Figdir}{.}

\newcommand{\myfiglabel}[1]{\label{#1}}
\newcommand{\comment}[1]{}
\newcommand{\bh}{\Bee\ hadron}

\def\hugehead{
\let\thanks=\footnote
} 

\newcommand{\sys }{\ensuremath{ _{\mathrm{sys }}}}
\newcommand{\stat}{\ensuremath{ _{\mathrm{stat}}}}
\newcommand{\institute}[1]  {}

\begin{document}

\preprint{DESY 05-233}

%
%
%
\title{ \bf
Improved Measurement of the {\boldmath \bbar }  Production Cross Section    
in 920 GeV Fixed-Target Proton-Nucleus Collisions  }

\author{
I.~Abt$^{23}$,
M.~Adams$^{10}$,
M.~Agari$^{13}$,
H.~Albrecht$^{12}$,
A.~Aleksandrov$^{29}$,
V.~Amaral$^{8}$,
A.~Amorim$^{8}$,
S.~J.~Aplin$^{12}$,
V.~Aushev$^{16}$,
Y.~Bagaturia$^{12,36}$
% ...

\vspace{5mm}

% ...
$^{8}${\it LIP Coimbra, P-3004-516 Coimbra,  Portugal~$^{f}$} \\
% ...
$^{10}${\it Institut f\"ur Physik, Universit\"at Dortmund, D-44221 Dortmund, Germany~$^{d}$} \\
% ...
$^{12}${\it DESY, D-22603 Hamburg, Germany} \\
$^{13}${\it Max-Planck-Institut f\"ur Kernphysik, D-69117 Heidelberg, Germany~$^{d}$} \\
% ...
$^{16}${\it Institute for Nuclear Research, Ukrainian Academy of Science, 03680 Kiev, Ukraine~$^{h}$} \\
% ...
$^{23}${\it Max-Planck-Institut f\"ur Physik, Werner-Heisenberg-Institut, D-80805 M\"unchen, Germany~$^{d}$} \\
% ...
$^{29}${\it DESY, D-15738 Zeuthen, Germany} \\
% ...
$^{36}${\it visitor from High Energy Physics Institute, 380086 Tbilisi, Georgia} \\
% ...
$^{d}$ supported by the Bundesministerium f\"ur Bildung und Forschung, FRG, under contract numbers 05-7BU35I, 05-7DO55P, 05-HB1HRA, 05-HB1KHA, 05-HB1PEA, 05-HB1PSA, 05-HB1VHA, 05-HB9HRA, 05-7HD15I, 05-7MP25I, 05-7SI75I \\
% ...
$^{f}$ supported by the Portuguese Funda\c c\~ao para a Ci\^encia e Tecnologia under the program POCTI \\
% ...
$^{h}$ supported by the National Academy of Science and the Ministry of Education and Science of Ukraine \\
% ...

}

\collaboration{ \bf The \hb Collaboration}
\noaffiliation
%

%
\vskip 1.0 cm
\begin{abstract}
A new measurement of the \bbar\ production cross section in 
920 GeV proton-nucleus collisions is presented by the \hb collaboration.
The \bbar\ production is tagged via inclusive bottom quark decays into
\jpsi\ mesons, by exploiting the longitudinal separation of \jpsill\ decay 
vertices from the primary proton-nucleus interaction point. Both \ee\ and \mm\
channels are reconstructed, for a total of $83 \pm 12$ inclusive \bjpsiX\
events found. The combined analysis yields a \bbar\ to prompt \jpsi\ cross
section ratio of
${\frac{{\Delta\sigbbar}}{{ \Delta\sigma_{\jpsi}}} }
= 0.032\pm 0.005 \stat \pm 0.004 \sys$ 
measured in the \xf\ acceptance ($-0.35 <\xf <0.15 $), extrapolated to
$\sigbbar = 14.9 \pm 2.2 \stat \pm 2.4 \sys \ \textrm{nb/nucleon}$ in the
total phase space.
\end{abstract}
\pacs{ \\
{13.85.Ni}{~Inclusive production with identified hadrons}\\
{13.85.Qk}{~Inclusive production with identified leptons, photons, or other
  non-hadronic particles }\\
{13.20.He}{~Decays of bottom mesons }\\
{24.85.+p}{~Quarks, gluons, and QCD in nuclei and nuclear processes}\\[2cm]
     } 
\maketitle

\section{Introduction}

Recent improvements in the theoretical description of heavy flavor ($c \bar
c$, \bbar )
hadroproduction, inspired by the availability of increasing amount of data
in various kinematic regimes, have provided new insight into the physics
of hadron-hadron collisions~\cite{cacc2004,frixioatDIS}.
Investigations in this field are further stimulated by the
current and the next generation of heavy ion colliders (RHIC, LHC) where
signatures of Quark Gluon Plasma (QGP) can be identified only after
hadron-hadron collisions in a non-QGP regime are understood with
sufficient accuracy.

In the field of fixed target bottom production, several experiments have been
performed over the past years at Fermilab, CERN and DESY with both pion and
proton beams (see \cite{nason1997,salam2002,lour2004} for recent reviews).
Several theoretical calculations have become available 
\cite{nason1998,vogt2004}, 
although the accuracy of the predictions is usually quite poor. 
The large
experimental and theoretical uncertainties provide a strong  motivation for a
more precise and accurate measurement of near-threshold \bbar\  production,
where the sensitivity to theoretical models is large and the
kinematic region is complementary to that of \ccbar\  or \bbar\
production at very high energies at collider experiments. In fact, in the
near-threshold region, perturbative QCD can be applied and  full
next-to-leading-order calculations are available, although
higher order terms are important and not yet fully evaluated.
In this regime, measurements have been published by only three experiments:
E789 \cite{E789bb}, E771 \cite{E771bb} and \hb
\cite{HBbb}.

The \hb experiment can measure the \bbar\  cross section via the
inclusive \bjpsiX\ decay mode. Interactions are produced by 
920 GeV protons in the halo of the HERA beam
impinging on target wires of different
materials at $\sqrt{s}= 41.6 \egev$ proton-nucleon 
center-of-mass energy.
 The \bbar\ production cross section (\sigB) on a nucleus of
atomic number $A$ is obtained via the inclusive reaction
\begin{equation}
pA \rightarrow \bbar \; X  \ \ \textrm{with} \ \ \bbar \rightarrow 
\jpsi \, Y \rightarrow  (e^+ e^- / \mu^+ \mu^-) Y .
\nonumber
\end{equation}
The \bh s decaying into \jpsi\ (``\bjpsi " in the following) are
distinguished from the large background of \jpsi\ mesons produced directly 
on the target (``prompt \jpsi " in the following) by exploiting the 
\Bee\ lifetime in a detached vertex analysis.
\hb is capable of detecting  both dilepton
decay channels of the \jpsi, providing an increase of the statistical
significance of the measurement and internal consistency checks.

A first measurement based on limited statistics was published in
\cite{HBbb}. This work presents an improved measurement based on
larger statistics collected during the 2002-2003 data taking. 
The larger sample of \bbar\ events allows us
to perform further cross checks,
 such as a lifetime measurement,
a search for specific decay channels of $B$ mesons
with a \jpsi\ in the final state and a search for inclusive decays of
the form $\bjpsi +h^\pm+X$.

The present paper is structured as follows. 
In Sect.~\ref{sec:meth} the basic concepts of the \sigbbar\ measurement
are discussed. An overview of the detector, the trigger performance
and the data samples is given in Sect. \ref{sec:det}, 
followed by a description of the Monte Carlo simulation  
(Sect.~\ref{sec:mc}).
The methods used for the \jpsi\ reconstruction and selection are presented 
in Sect.~\ref{sec:sel} and discussed separately for the
muon (\jpsimm) and the electron (\jpsiee) channel.
Sect.~\ref{sec:anal} describes
the procedure for the identification of secondary vertices from
 \bjpsi\ decays and their separation from the prompt \jpsi\
background. 
In Sect.~\ref{sec:btag}, independent confirmations of the \Bee\  content
of the selected events are provided.
In Sect.~\ref{sec:comb}, the muon and
electron data are combined into the final result, which is then 
compared  with the existing experimental data on \sigbbar\  
and recent theoretical predictions. The main systematic 
uncertainties of the measurement are discussed. 
Sect.~\ref{sec:concl} provides a brief summary of the analysis.


\section{Measurement Method}                       \label{sec:meth}
The total \bbar\ production cross section in proton-nucleon ($p$N)
interactions, integrated over the complete phase space, is the
quantity of interest for characterizing the \Bee\ production rate
since it is most readily compared to theoretical calculations and
other experimental results. However, in fixed-target experiments, 
the proton-nucleus ($p$A) cross section 
is measured for the actual detector acceptance (\dsigB ) and then 1) scaled to
the proton-nucleon cross section taking into account the
known (or assumed) nuclear dependence and 2) extrapolated
to the full phase space by using theoretical models or information from other
experiments. 

In the present analysis, systematic uncertainties related to
detector and trigger efficiencies are minimized by measuring the \bbar\  cross
section relative to the \jpsi\ cross section in the kinematic region
in our acceptance.
The analysis consists essentially of two steps:
1) the \jpsi\ selection to determine the number (\nP ) of prompt \jpsi\ mesons
(Sect.~\ref{sec:sel}) which are copiously produced on the target;
2) the selection and counting of \jpsi\ mesons 
  detached from the primary interaction vertex 
(Sect.~\ref{sec:anal}), originating from \bjpsi\ decays  (\nB ). 
Following this strategy, the ratio  of
\bbar\  to prompt \jpsi\ production cross sections 
in our acceptance (\dsigB/\dsigP )  
can be written as
\begin{equation}
\frac{\dsigB}{\dsigP} = \frac{\nB}{\nP} \cdot \frac{1}{ 
  \effR \cdot 
  \effBz \cdot \Br{\bbjX}}, 
\label{eq:ds}
\end{equation}
with $\effR={\effB}/{\effP}$, where
 \effB\ and \effP\ are the detection efficiencies
(including trigger, reconstruction and selection)
for \jpsi\ from \Bee\ decays and prompt \jpsi,
respectively. \effBz\ is the detached vertex selection 
efficiency. $\Br{\bbjX} = 2 \cdot (1.16\pm 0.10)\%$~\cite{PDG}
is the inclusive 
$b \bar b \to \jpsi X$ branching ratio, assumed to be the
same in hadroproduction as that measured in $Z$ decays.

All the quantities entering into Eq.~(\ref{eq:ds}) 
(except the branching ratio) are
evaluated for \jpsi\  mesons produced in the 
kinematic range covered by \hbp. In terms of
 Feynman-$x$ (\xf ) and transverse momentum (\pt ),
the kinematic range covers the intervals $-0.35 < \xf <0.15$
and $0<\pt <6 \ \pgev$. This will be indicated in the following as the 
\hb\ acceptance range.

The prompt \jpsi\ cross section (\sigP) has been measured 
at various energies ($\sqrt{s} \in [10,\, 200] \egev$) 
and using various target materials ($A \in [1, \, 197]$)
both in fixed target and collider experiments.
\sigP\ is usually expressed
by parameterizing the atomic weight ($A$) dependence of the cross section as
$\sigP=\sigma_{\jpsi} \cdot A^\alpha$, where 
$\sigma_{\jpsi}$ is the \jpsi\ production cross section for proton-nucleon
interactions and $\alpha$ has been
measured with high statistical precision as a function of \xf\ and \pt\ by the 
E866 collaboration \cite{E866} providing an average value of $\alpha=0.96 \pm
0.01$ in our \xf\ range.

The actual value of $\sigma_{\jpsi}$ is needed to obtain
the absolute $\sigma(\bbar)$ cross section and it is known \cite{cacc2004} 
that the values at energies around $\sqrt{s} = 20-50$ \egev\ scatter
strongly. To determine the value of $\sigma_{\jpsi}$ at the \hb energy,
a global analysis~\cite{refjpsi} has been performed on all published \jpsi\
cross section measurements\footnote{The global fit includes our own
  measurement of the \jpsi\ cross section ($663 \pm 74\pm 46$ nb/nucleon
  \cite{hbjp}). We prefer to normalize the present measurement to the value
  given by the global fit since the fit provides a complete summary of all
  available measurements.}. The best value, obtained from a fit on 
$\sigma_{\jpsi}(\sqrt{s})$ data with the help of a NRQCD model 
evaluated at the next-to-leading order is
$\sigma_{\jpsi}(41.6 \egev) = (502 \pm 44)\ \mathrm{nb/nucleon}$.
The main uncertainty is systematic and has been evaluated by changing 
the selection of
data submitted to the fit and by changing the free parameters in the model.
Taking into account that the fraction of prompt \jpsi\ mesons  produced in the 
 \hb\ acceptance range
 is $\fP\ =(83 \pm 1)\% $ (see
Sect.~\ref{sec:mc}), 
we will assume a reference
prompt \jpsi\ cross section in our  acceptance which is given by:
$\Delta\sigma_{\jpsi} = \fP\ \times \sigma_{\jpsi} =
(417 \pm 37 )\ \mathrm{nb/nucleon}$.

The $A$ dependence of the \sigB\ cross section has not been measured. 
In this paper, we assume the form 
$ \sigB\ = \sigma(\bbar) \cdot A^{\alpha_\Bee}$,
with $\alpha_\Bee =1$ since no nuclear suppression is expected 
in open bottom or open charm hadroproduction \cite {Adep}. This expectation is
confirmed in $D$-meson production data \cite{Ddep}.  
The same assumption is made in all published fixed-target $\sigma(\bbar )$
measurements. 

The different $A$ dependences of prompt \jpsi\ and \bbar\
cross sections complicate the evaluation 
of the cross section ratio when combining data sets 
from different target materials and several
approaches are possible. Since in the present case \nB\  suffers
from low statistics in some sub-samples,
we express the  \bbar\ to \jpsi\ cross section ratio per nucleon in our
acceptance range ($R_{\Delta  \sigma}$) as:

\begin{eqnarray}
&&R_{\Delta \sigma} = \frac{\dsigbbar}{\Delta\sigma_{\jpsi}} = \nonumber \\
&&=  \frac{\nB }{ \Br{\bbjX} \cdot \Sigma_i n^i _P \cdot
  \eff_{R, i} \cdot  
 \eff_{\bbar , i}^{\Delta z}  \cdot A_i ^{\alpha_\Bee-\alpha}},\qquad 
\label{eq:dsigbbarss}
\end{eqnarray}

\noindent
where \nB\ is the total number of detached \jpsi\ events found in the 
acceptance, $n^i _P$,
$\eff_{R, i}$ and $\eff_{\bbar , i}^{\Delta z}$ are respectively the numbers of
prompt \jpsi , the \jpsi\ selection efficiency ratios and the detached
vertex selection efficiencies measured on the sub-sample $i$ taken with a
target of atomic weight $A_i$. 
The quantity defined by Eq.~(\ref{eq:dsigbbarss}) has the 
advantage of minimizing the dependence on theoretical models and on
extrapolations.
We also quote the \bbar\
cross section for full phase space, evaluated as $\sigbbar = R_{\Delta
\sigma} \cdot \Delta\sigma_{\jpsi} / \fB$, where \fB\ is the fraction of
\jpsi\ mesons from \Bee\ decays in the \hb\ acceptance range. 
The quantity \fB\ is
determined through theoretical models as discussed in Sect.~\ref{sec:mc}.


\section{Detector, Trigger and Data Sample}        \label{sec:det}

\hb \cite{hb_tdr,hb_oct} 
is a fixed target experiment at the HERA storage ring at DESY.
The spectrometer and the trigger system were designed for efficient
real-time filtering of \bh s based on an online reconstruction of \jpsi\
mesons.

\subsection{ Detector}

\begin{figure*}
\resizebox{0.95\textwidth}{!}{%
  \includegraphics{\Figdir/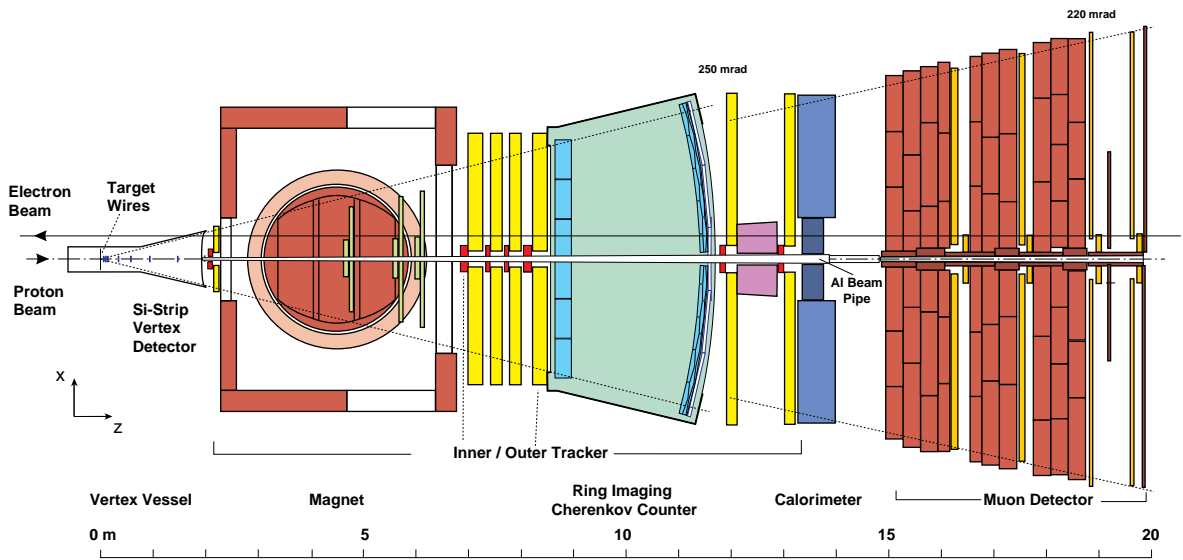}
}
\caption{Plan view of the \hb detector.}
\myfiglabel{fig:detect}       
\end{figure*}

The spectrometer (Fig.~\ref{fig:detect})
has a forward geometry, covering from 15 
to 220 mrad in the bending plane, and from 15 to 160 mrad 
vertically.
The detector features high resolution
tracking and vertex reconstruction, good particle identification
over a large momentum range, and a multilevel dilepton trigger.

The wire target \cite{hb_targ} consists of two independent
stations containing 4 wires each and separated by 40 mm along the
beam direction. The wires are made of various
 materials  (C, Al, Ti, Pd, W) with dimensions
 of 50--100 $\mu$m perpendicular to
the beam and 50--500 $\mu$m along the beam. 
Up to eight target wires can be 
operated simultaneously and each wire can be moved independently
into the proton beam halo to adjust the interaction rate. 
During the data taking, the target operated at rates of about 7 MHz, both in
single and double wire configurations. 

The Vertex Detector System (VDS) \cite{hb_vdet}, 
located downstream of the target, consists
of 8 planar stations with a total of 64 silicon microstrip detectors 
($\approx 50 \mu$m readout pitch).
It provides high spatial resolution 
for the reconstruction of primary and secondary vertices. A resolution
of about 450 $\mu$m on \jpsi\ decay vertices along the direction of flight
was achieved, fulfilling design specifications.

The main tracking system is located behind a 2.13~T$\cdot$m bending
magnet and extends to 13 m downstream of the interaction region. 
The granularity of the tracking system was adapted to the increasing particle 
density with decreasing distance to the beam.
The Inner TRacker (ITR) \cite{hb_msgc} covers the range 
up to 20 mrad with Micro Strip Gas Chambers (MSGC) equipped with GEM foils.
The Outer TRacker (OTR) \cite{hb_otr}, made of  honeycomb
drift chambers, reached a hit resolution of about 350 $\mu$m.
The overall tracking system, consisting of VDS, OTR and ITR, reached a
muon momentum resolution of $\Delta p/p (\%) = (1.61\pm 0.02) \oplus (0.0051
\pm 0.0006) p$ where the momentum $p$ is in \pgev\ and in the 
range $5<p<80 \, \pgev$.

Particle identification is performed by a Ring Imaging Cherenkov 
detector (RICH), 
an Electromagnetic CALorimeter (ECAL) and a MUON
system.

The RICH detector~\cite{hb_rich}
provides pion--kaon--proton separation.
It uses a $2.5~\mbox{m}$ long $\mbox{C}_{4}\mbox{F}_{10}$ radiator volume
for Cherenkov
light emission by charged particles, which after reflection
in spherical and planar mirrors, is directed to one of two planes of 
photomultiplier tubes.

The ECAL~\cite{hb_ecal} is optimized for good
electron/gamma energy resolution and electron-hadron discrimination.
It is composed of 5956 independent calorimeter
modules built with shashlik technology. The cell size increases 
with the distance from the beam.
The detector is instrumented with a fast digital read-out and a pretrigger
system which provides reconstructed clusters with large transverse energy 
(pretrigger seeds). 
The energy resolutions for the three ECAL regions (Inner, Middle, Outer) are,
 respectively, $\sigma_E / E =
20.5 \% /\sqrt{E}~\oplus 1.2\%$,
$\sigma_E / E = 11.8\% /\sqrt{E}~\oplus 1.4\%$  and $\sigma_E / E = 10.8\%
/\sqrt{E}~\oplus 1.0\%$ \cite{bgsiena}.
Spatial resolutions, determined with $\pi^0$ decays,
are ranging from 
1 to 8 mm
depending on the 
ECAL region and the particle energy.

The MUON system~\cite{hb_mu} consists of four tracking stations located
in the most downstream part of the detector at various depths in an
iron and concrete muon--absorber. The active elements are mainly 
conventional tube chambers arranged in double layers. Pad cathodes coupled to
the tubes of the two most downstream tracking stations are used for
pre-triggering purposes. 

\subsection{ Trigger Configuration}

The dilepton trigger is initiated by pretrigger (PT) signals from either
the MUON system~\cite{hb_mu_pre} or the ECAL~\cite{hb_ecal_pre}. 
Dual pad coincidences in the MUON system or complete ECAL
clusters with a transverse energy above 1 GeV constitute the basic
trigger seeds. These are used by the First Level Trigger (FLT) ~\cite{hb_flt}
to define a sequence of regions of
interest inside the OTR, which are used
to initiate a search for track
candidates originating from \jpsi\ decays.
Track parameters as well as the PT seeds are then sent to the 
Second Level Trigger (SLT)~\cite{hb_slt},
which is a highly configurable software filter that
extra\-po\-lates found tracks through the magnet and VDS, finally applying
a mild vertex constraint.

More specifically, for the data presented here, the FLT required the presence
of at least two PT seeds and at least one reconstructed track originating from
one of the PT seeds. The SLT searched for tracks starting from the PT seeds,
requiring that at least two complete tracks, with segments in the OTR and the
VDS, be found and that the two tracks be consistent with a common vertex
hypothesis. Moreover, the SLT track-seeding algorithm imposed a target 
constraint by including in the initial track parameters a point near the
center of the active target wires with uncertainties large enough to
accommodate tracks from all targets and also to allow for the finite $B$ 
lifetime.
Accepted events were read out and sent to an online reconstruction
farm~\cite{hb_daq}.

The \jpsi\ trigger worked constantly at rejection factors
around $4 \times 10^4$, matching the typical data recording rate
of 120 Hz (to tape).  
The experiment routinely achieved an 
event yield of about 1200 \jpsi\ mesons per hour at an
interaction rate of about $7~\mbox{MHz}$.

\subsection{ Data Sample } \label{sec:sub-sample}

The data sample used in this analysis was collected between
October 2002 and March 2003.
During this period, 164 million dilepton-triggered events were 
recorded containing about 300,000 \jpsi\ mesons.
These are distributed almost equally between the
dimuon and dielectron channels.

Data were taken in nine different wire configurations
of single or double wire runs, for a total of 14 equivalent single-wire
samples. The wire materials used were carbon
(A=12, $\approx$64\% of the \jpsi\ sample), tungsten (A=184, $\approx$27\%),
and titanium (A=48, $\approx$9\%).

During the data taking period, the detector was operated
under constant monitoring. The data quality
was assessed both online and offline. Only runs with good performance 
of all the main detector components and stable trigger conditions are
used. 
Five periods with constant experimental conditions 
were defined and the data were grouped accordingly.


\section{Monte Carlo Simulation}                   \label{sec:mc}

Monte Carlo (MC) simulations are used to determine the
efficiency terms entering into the cross section
equation (Eq.~\ref{eq:dsigbbarss}), to determine the
criteria to select the \bjpsi\ candidates, and 
to understand the nature of the background.
The simulation includes our best knowledge of the physics of the processes
under investigation, of the detector, of the trigger and of the event 
reconstruction.

To best describe interactions in the \hb\ environment, 
the Monte Carlo generator exploits a combination of features of different
standard tools. 
The simulation of heavy quark ($Q$) production is obtained
by generating the basic process $pN \to Q \bar Q X$ 
with {\sc Pythia 5.7}~\cite{pythia} and 
hadronizing the heavy quarks using the {\sc Jetset 7.4}~\cite{pythia} package.
In the second step, the remaining energy and momentum of the collision  
is used by
{\sc Fritiof 7.02}~\cite{fritiof} to provide the underlying inelastic event and
to generate further interactions 
inside the target nucleus. After these steps, the generated particles
are passed through the {\sc Geant 3.21} package \cite{geant} for 
detector simulation.

For an accurate description of the kinematic characteristics of prompt \jpsi\
production, experimental data on prompt \jpsi\ differential cross
sections $d \sigma /d(p_T^2)$  and $d \sigma /d x_F$ \cite{E771dj} are
used to tune the standard generator: all the produced events are weighted
to match the available proton-silicon measurements on 
differential \jpsi\ cross sections. 
Since these are available in the positive \xf\
region only, we assume a symmetric \xf\ distribution 
of prompt \jpsi\ production: 
$d \sigma /d x_F = \beta \cdot (1-|\xf |)^c$ where $c=6.38\pm 0.24$
\cite{E771dj}. 
This extension conforms both to all the basic charmonium
production models and to our own data on the \jpsi\  differential cross
sections \cite{hb_jpdiff}.
The model dependence of the generated \pt\ spectrum is of minor importance
since the acceptance for the \jpsi\ \pt\ is essentially flat. 
The influence of a possible \jpsi\ polarization within the limits given
by experimental results~\cite{E771dj, pola} 
is taken into account in the systematic
uncertainties of the prompt \jpsi\  MC.
With this model, the fraction $f_p$ of prompt \jpsi\ mesons  produced in the 
\hb\ acceptance range
is $f_p = (83\pm 1) \%$, 
where the error is due to the uncertainty in the exponent $c$ of the
proton-silicon data fit \cite{E771dj}.

Since, for open \Bee\  production, there are no measured differential cross
sections, we use a \bbar\  production model which is
mostly based on NRQCD theory \cite{mang}. 
As in the case of charmonium production, 
the {\sc Pythia} generated \bbar\  events are weighted with a model containing
several contributions:
1)~the generated \Bee\ quark kinematics (\xf\ and \pt ),
as given by the computation of Mangano {\it et al.}, using 
the most recent next-to-leading-order 
MRST parton distribution function \cite{mrst} with a \Bee\ quark mass of 
$m_{\Bee} = 4.75$~\mgev\  and a QCD renormalization scale 
$\mu=\sqrt{m^2_\Bee + \pt^2}$;
2) the intrinsic
transverse momenta  of the colliding quarks, smeared with a Gaussian
distribution resulting in $\langle k^2 _T \rangle = 0.5 \ \egev^2/c^2$
\cite{chay}; and
3) the \Bee\ fragmentation function,
described by a Peterson shape \cite{peterson} with a parameter 
$\epsilon = 0.006$ \cite{epsil}. 
The subsequent \bh\  formation and decay are described by the 
{\sc Pythia} default parameters. Possible \bh\ 
polarization and \jpsi\ polarization in \Bee\ decays are assumed to be null.
With these assumptions, the fraction of detached \jpsi\ mesons  produced 
in the \hb\ acceptance range (see Sect.~\ref{sec:meth})
is $\fB\ = (90.6\pm 0.5) \%$, where the error is determined in the systematic
studies described below.

The sensitivity of the measured cross section
ratio $R_{\Delta \sigma}$ to the \bbar\ MC model parameters 
has been determined by varying:
the parton distribution functions (from MRST to CTEQ5 \cite{cteq5}),
the \Bee\ quark mass (in the range
$m_\Bee \in [4.5, 5.0]$~\mgev), the QCD renormalization scale (from 
$0.5\sqrt{m^2_\Bee + \pt^2}$ to $2\sqrt{m^2_\Bee + \pt^2}$), the fragmentation
function (from the Peterson form~\cite{epsil,E789bb,aleph} with parameter
$\epsilon \in [0.002, 0.008]$, to the Kartvelishvili form \cite{kart}
with parameter $\alpha_\beta = 13.7 \pm 1.3$ \cite{aleph}), the 
intrinsic transverse momentum distribution 
(with $\langle k^2 _T \rangle$ in the range $[0.125, 2.0]$ $\egev^2/c^2$) and
the fraction of \Bee\ baryons produced in the \Bee\  hadronization process 
in the range $[0, 12]$\%. 
The observed variations in the detection efficiencies have been included in the
systematic uncertainty. 
No significant dependence ($< 1.5 \%$) of the cross section ratio has 
been found on the momentum of the \jpsi\ mesons in the \bh\   rest frame nor
on the polarization of \jpsi\ mesons coming from \Bee\ decays.

The {\sc Geant} tracking step is performed with the help of 
a detailed description of the detector which includes 
both the active (instrumented) and inactive (support structure) elements.
The detector response is simulated by reproducing
the digitization of electronic signals and of the read-out chain.
A status (masked/dead/alive) and an efficiency is associated to each channel
to take into account the measured efficiency in data as well as the
faulty/noisy/masked channels. This information was produced for each 
data-taking sub-period (see Sect.~\ref{sec:sub-sample}).  
The MC events are subjected to a full trigger simulation and reconstructed
with the same code as the data. 
For each data sub-sample, 
MC samples were produced for the signal channels and background sources 
to evaluate the efficiency terms in Eq.~(\ref{eq:dsigbbarss}) for
the specific running conditions.

\section{ \jpsi\  Event Selection}        \label{sec:sel}

The first stage of the analysis is devoted to the determination of the
number of \jpsi\ mesons produced at the target,
which is dominated by the prompt production.
This is done in several steps: selection of well-measured tracks,
lepton identification, lepton pair selection and \jpsi\ candidate counting.
Most of the selection criteria are common for the muon and the electron 
final states and are described first. Other details specific to the different
lepton types (for example, particle identification
and counting of prompt candidates) are discussed separately.

The cuts applied for prompt  \jpsi\ signal selection mostly reflect the 
detector acceptance and requirements already explicitly or implicitly 
imposed by the trigger algorithms.  Therefore the lepton track search
initially uses information provided by the trigger.
The parameters of the SLT track candidates 
that resulted in a trigger decision are given as seeds to the
offline reconstruction program. If a seed successfully results in 
a reconstructed track, it is flagged as a 
`trigger track'. Only these
are considered in the search for \jpsi\ candidates.

Further selection is performed
on the offline reconstructed tracks, which profit from
improved reconstruction due to more accurate 
calibration and alignment constants compared to those used online. 
Basic cuts on the number of hits in
the VDS and in the main tracker (OTR+ITR), and on the $\chi^{2}$ probability 
of the track fit are used to 
select only well-reconstructed tracks and to reject mismeasured and ghost
tracks.
Weak cuts on momentum ($6.0 < p < 200~\pgev$) and transverse momentum 
($0.7 < \pt < 5.0~\pgev$) are applied.
The lower cuts mostly reflect the detector acceptance and the trigger
requirements. In the muon case, they also
improve the rejection of
hadrons traversing the absorber and of muons from $\pi/K$
decays in flight. The upper cuts 
discard tracks with momenta above the expected range
for \jpsi\ decay products.

To complete the single lepton selection, requirements 
on the relevant particle identification specific to muon or electron channels are
imposed and are discussed
separately in Sect.~\ref{promptmm} and \ref{promptee}.

When two lepton candidates with opposite charge are selected, 
a vertex fit is performed. Only dilepton candidates with
a $\chi^{2}$ probability greater than $1\%$ are considered further. 
This requirement ensures that the tracks 
originate from the same vertex and helps to reject combinatorial
background and background from  
\ccbar\ and \bbar\ double semileptonic decays in the subsequent 
detached vertex analysis.
Moreover, the dilepton \xf\ is required to be between $-0.35$ and 0.15
to match the detector acceptance.
Only one dilepton candidate per event is accepted. When more than one
dilepton
candidate survives all cuts, the one with the best particle identification 
for both leptons is chosen. 

\subsection{  \jpsimm\ }
\label{promptmm}

The identification of muon tracks is simplified by the fact that
only muons have a significant probability of penetrating through the
absorbers of the MUON detector. Therefore a minimal set of cuts
on signals from this detector (hit multiplicity and muon likelihood
${\cal L}_{\mu}$) is sufficient for robust muon 
identification. 
In case more than one muon pair is selected,
the pair with the highest product of muon likelihoods is chosen.

The dimuon mass spectrum obtained with this selection 
is shown in Fig.~\ref{fig:JPSITRIGPEAK}a.
The number of prompt \jpsi\ candidates is determined by fitting the
mass distribution to a function derived mainly from Monte Carlo studies~\cite{spirid}.
This function includes: 1) a description of the non-Gaussian shape of the
\jpsi\ signal\footnote{
A symmetric function composed of three Gaussians is used, to take
properly into account the details of the momentum resolution in
a complex apparatus such as \hbp.},
2) a radiative tail due to the $\jpsimm \gamma$ 
decay, 3) a Gaussian shape for the $\psi(2S)$ signal and 4) an exponential
function for the background. 
 
The mean value of the \jpsi\ mass is 
$3093.5 \pm 0.2_\mathrm{stat}$ MeV/$c^2$, and
the full width at half maximum (FWHM) of the background subtracted \jpsi\
peak is 90 MeV/$c^2$. The shift observed in the \jpsi\ mass with respect to
the PDG value~\cite{PDG} is due to small systematic effects in the tracking
and in the internal alignment of the detector.
The total number of reconstructed \jpsi\ mesons is $\nP (\mm ) = 148200 \pm
500$. 
This number depends slightly on the fitting function and mass interval
considered. These uncertainties are included in the
systematic error (Sect.~\ref{sec:comb}).

\subsection{  \jpsiee\ }
\label{promptee}

The electron and positron selection is
affected by larger background contributions than the muon case, 
mostly due to pions interacting in the ECAL and hadrons overlapping with
energetic neutral showers. 
Requirements on the lepton identification are therefore more demanding than
in the muon case.
Additional corrections are used to improve the accuracy of the
electron momentum estimate.

The electron track candidates are selected by requiring
an ECAL cluster with a transverse energy greater than $1.0~\egev$,
matched to the track, both at the trigger and the offline analysis levels.
In the non-bending direction ($y$), the distance $\Delta y$ between
 the track position extrapolated to the ECAL
and the reconstructed ECAL cluster must satisfy the requirement 
$|\Delta y/\sigma_{\Delta y}|<3.0$, where $\sigma_{\Delta y}$ is the 
resolution on $\Delta y$ for clusters and tracks from $e^\pm$.
The typical resolutions are in the range $3-10$ mm depending on the ECAL
region.  
Cluster-track matches from random combinations and from hadron interactions in
the calorimeter are substantially suppressed by this cut since their $\Delta y$
distributions are wider by more than a factor two compared to $e^\pm$
tracks. 

For a precise measurement of the electron momentum, 
some corrections to the value provided by the tracking system 
are needed.
During their passage through the material in front of the ECAL,
electrons can emit bremsstrahlung (BR) photons and lose energy due to
ionization 
processes. ECAL clusters due to BR photons emitted before the 
magnet are searched for in the ECAL 
at a location given by the extrapolation of the initial track direction to the
ECAL position. If a BR cluster  
with a measured energy $E_\mathrm{BR} > 1~\egev$ is found, the energy of the cluster is added
to the electron momentum at the production point.
The correction ($\approx 1-2\%$) for further energy losses is evaluated
stochastically via Monte Carlo simulations.

Electron identification in the ECAL is mainly based on the ratio of the 
cluster energy ($E$) to the momentum obtained from the tracking ($p$). 
Typical $E/p$ distributions for $e^{\pm}$
 signals have a Gaussian shape with
mean $0.99$ and width $\sigma\approx 6\%$.
Since the requirement on $E/p$ 
is also crucial in the selection of the detached signal,
it was included in the optimization procedure described in
Sect.~\ref{sec:anal}. 
The obtained cut values are $-2.0~\sigma < E/p-0.99 < 3.5~\sigma$. If more than
 one electron-positron pair is found in one event, the combination with the
 best 
 $E/p$ ratio is taken.

Requiring that at least one of the electrons from a \jpsi\ candidate has
an associated bremsstrahlung photon (BR tag) or that both leptons
have BR tags increases the signal-to-background ratio from 0.6 to 3
and 9, respectively, providing a powerful method to isolate clean
samples. These are used for studies of resolutions (for example, of $\Delta y$
and $E/p$) and for cross checks.
As explained in~\cite{HBbb}, the probability that a BR photon is emitted
and detected in a separated ECAL cluster (bremsstrahlung tag probability) 
can be measured redundantly by counting \jpsi\ mesons for different
tag criteria. For the current setup, 
the average tag efficiency ($\varepsilon_\mathrm{BR}$) 
is $\varepsilon_\mathrm{BR} = 0.30 \pm 0.01\stat \pm 0.01 \sys$ per track. 
The same analysis, performed on MC
events, yields a MC tag probability of $\varepsilon_\mathrm{BR}^{MC} = 0.29 \pm
0.01\stat \pm 0.01\sys$, in agreement with data. 
This gives further confidence in the quality of our
detector description and MC simulation, in particular in the crucial region
in front of the magnet, where the triggered electrons cross about 0.07
radiation lengths of material.

The BR tag played a central role in the previous analysis \cite{HBbb}, since
it was used for counting the number of prompt \jpsi\
candidates. Subsequent improvements in ECAL stability and removal of material
in the magnet region resulted in reduced combinatoric 
background and better mass resolution. Consequently the BR tag is now
only used for calibration and cross-checks and not for signal
counting.

The invariant mass spectrum obtained for the
final \jpsiee\ sample is shown in Fig.~\ref{fig:JPSITRIGPEAK}b.
As for the muon case, the number of prompt \jpsi\ candidates is determined 
by fitting the mass distribution with a function determined from MC studies.
It includes 1) Gaussian functions for the \jpsi\  and for the 
$\psi(2S)$ signal, 2) a function to
take into account the radiative tail of the charmonium decays and
3) an exponential-like function for the background.
The radiative tail contains both the partially reconstructed
\jpsiee $\gamma$ decays as well as the tail
due to incomplete electron energy measurement or BR recovery.
The mean value of the \jpsi\ mass is $3109 \pm 1_\mathrm{stat}$ MeV/$c^2$, and
the FWHM of the background subtracted \jpsi\
peak is 140 MeV/$c^2$. 
The total number of \jpsiee\ candidates is
$\nP (\ee ) = 103800 \pm 1000$.

\begin{figure}[htb]
  \centering    
  \epsfig{file=\Figdir/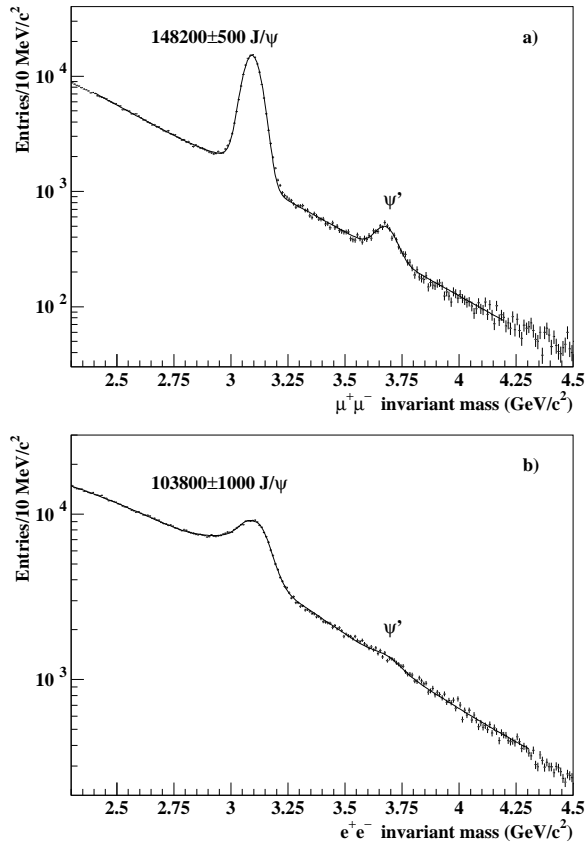,width=0.48\textwidth}
  \caption{
    Invariant mass plots for both \jpsi --decay channels: a) muon sample, b)
    electron sample.
    }
  \myfiglabel{fig:JPSITRIGPEAK}
\end{figure}


\section{Detached Vertex Analysis}                 \label{sec:anal}
In the second stage of the analysis, \jpsi\ mesons 
produced in the decay of a \bh\ are searched for. 
The mean decay length of a \bh\ in the \hb acceptance 
is about 8 mm. Accordingly, the \jpsi\ mesons produced in \Bee\ decays are
typically well-detached from the primary interaction at the target wires.
The \bjpsi\ candidate selection is performed in two tightly correlated steps:
first, a \jpsi\ candidate (as described in the
previous section) is assigned to one target wire 
and to one primary interaction vertex. Then, if 
the association is unambiguous, cuts designed to distinguish 
\jpsi\ mesons coming from \Bee\ decays from those produced promptly
at the target are applied.
Almost all cuts are applied on the significance of the relevant quantity
(ratio of a quantity to its uncertainty) 
rather than on its unnormalized value, to take advantage of the error
information. The variables used to isolate detached \jpsi\ mesons
are described in the following.

The decay length is defined as the distance between the primary 
interaction point and the \jpsi\ vertex along the proton beam direction 
(which defines the \hb $z$ axis). The positions of target wires, averaged over
many consecutive events, define better the primary interaction point than 
the fitted positions of individual primary vertices; 
therefore the decay length is calculated with respect to the target wire. 
With a dilepton vertex resolution of about 450 $\mu$m in $z$, 
the decays of \bh s can be separated from primary vertices
with high efficiency.

For \jpsi\ mesons produced at a detached vertex, each of the decay leptons
is  likely to be inconsistent with being produced at the target.
Therefore a cut on their impact parameters to the wire, defined as
the lepton distance from the wire at the $z$-coordinate of the wire, 
is applied. This requirement is very effective for
reducing background, as the wire position is
known in two dimensions with high precision. 

On the other hand, since the direction of the produced \jpsi\ meson is nearly
collinear with that of the \bh , 
the impact parameter of the \jpsi\ to the wire is
typically small (mostly below $20\, \mu$m). 
An upper cut is applied to remove combinatorial background, 
as well as background from semileptonic \bbar\ 
and double semileptonic \ccbar\ decays.

For double wire runs and for multiple primary vertex events, the
quantities described above can only be defined after 
association of the dilepton candidate to the correct target wire 
and to the correct primary interaction vertex.
For this purpose, we define two $\chi^2$-like quantities based on 
the lepton and the dilepton impact parameters to a wire and on their 
distances of closest approach to a primary vertex. The wire
and the vertex for which the corresponding $\chi^2$ is minimum are associated
to the dilepton candidate. 
Events in which the wire of the dilepton and the wire of the 
associated primary vertex do not match are discarded.
In this way, the probability for wrong wire assignment, estimated from 
\bbar\ MC events, is less than $1\%$. 

Moreover, in the case of double wire runs, it is possible that a lepton
candidate from an interaction
on one wire is combined with a lepton candidate originating from the other
wire. Since the reconstructed vertex of these two leptons tends to lie
between the two wires, these events are a dangerous source of background for
the detached signal. To reduce this type of background,
events in which one or both leptons are pointing to a reconstructed primary
vertex on
a different wire than the assigned one are discarded.

Events having unambiguous wire-vertex association are 
analyzed further. We exclude events with the
absolute value of the decay length 
$|\Delta z|< 2$ mm (fiducial volume cut), in order to 
ensure that the \jpsi\ vertex is well outside of the target wire.
This removes more than 99\% of prompt \jpsi\ background while keeping more
than 75\% of the signal.
Afterwards, on the total sample
 a cut optimization procedure is applied, on the following quantities:
the $\Delta z$ significance, the lepton and the dilepton
impact parameters to the wire and the lepton and dilepton distances of
closest approach to the primary vertex and $E/p$ for electrons. 

\subsection{  Cut Optimization Procedure}

To avoid biasing our signal yield and resulting
cross section ratio, we choose our selection criteria
'blindly', i.e. without looking at events in the signal
 region until all selection criteria are finalized.
The quantity to be maximized as a function of the cuts is the signal
significance ${S}/{\sqrt{S+BG}}$ in a fixed dilepton invariant mass
window, where the signal ($S$) is taken from \bjpsi\ MC events, and the
background ($BG$) is obtained by combining MC and real data.
The main background sources are listed below.
The first three contributions are estimated from MC, the fourth from data:

\begin{itemize}
\item A possible prompt \jpsi\ background surviving detachment cuts is one of
our main concerns, since there is no way to distinguish it from \jpsi\ mesons
coming from \Bee\ decays. Its yield is proportional to the prompt \jpsi\ cross
section. A large MC sample ($\approx 10^7$ events) is used to estimate the
surviving background contribution.
\item A large fraction of the 
background is coming from \bbar\ events
where both \Bee\ quarks produce a high-\pt\ lepton. 
The yield of this background is proportional
to the \bbar\ cross section.
\item Double semileptonic \ccbar\ decays can contribute in a similar way.
The intrinsic \pt\ cut at the trigger level removes most of this background. 
Nevertheless this contribution cannot be neglected,
since the cross section for \ccbar\ events is larger than the $b\bar{b}$ cross
section by three orders of magnitude.
\item Another contribution is due to combinatorial background, i.e.\
dileptons
made of incorrectly reconstructed or fake tracks, or tracks coming
from kaon or pion decays in flight. Its size can be estimated
from the number of events in which the dilepton vertex is reconstructed in the
unphysical region upstream of the target. Also, like-sign lepton pairs 
(available only for the muon channel)
or dileptons artificially formed by tracks from different events can give an
estimate of this background. 
\end{itemize}

The background used  for the cut optimization is an appropriately
scaled combination of
 prompt \jpsi\ MC, \bbar\ and \ccbar\ double semileptonic
MC events and either like-sign data events
(in the muon channel) or upstream data events for the  
combinatorial background.
The relative contribution of the different samples is estimated
on the basis of the relative cross sections (for the MC samples)
and from the upstream spectrum for the 
combinatorial background.
Since the  \bbar\ cross section is not known (being the goal of
the current measurement), its value 
is left to vary in a range between a quarter
and double of our previous measurement \cite{HBbb}. The
optimization is repeated for several intermediate values.  A stability
check is also performed by varying the initial values of the cuts in
order to verify the independence of the minimization results on
the initial conditions. 

This approach can work effectively only if the data distributions are well
reproduced by MC. This is verified 
for all the quantities used for the selections.
As an example, in
Fig.~\ref{fig:lepimp}, the distributions of the lepton impact parameter
to the wire are shown
for a sample of background subtracted \jpsimm\ (a) and a sample of background
subtracted double bremsstrahlung tagged \jpsiee\ (b). Agreement between
data and MC is obtained for all variables used in the optimization procedure
for both channels. 

Of the six parameters entering in the optimization procedure, 
the cuts on the lepton's and dilepton's closest distance of approach to the
primary vertex are fixed by the optimization procedure
to a value where all events surviving the other cuts are selected. 
Therefore these cuts have been removed from the optimization procedure 
and will not be mentioned further.

\begin{figure}
  \begin{center}
\resizebox{0.48\textwidth}{!}{%
\includegraphics{\Figdir/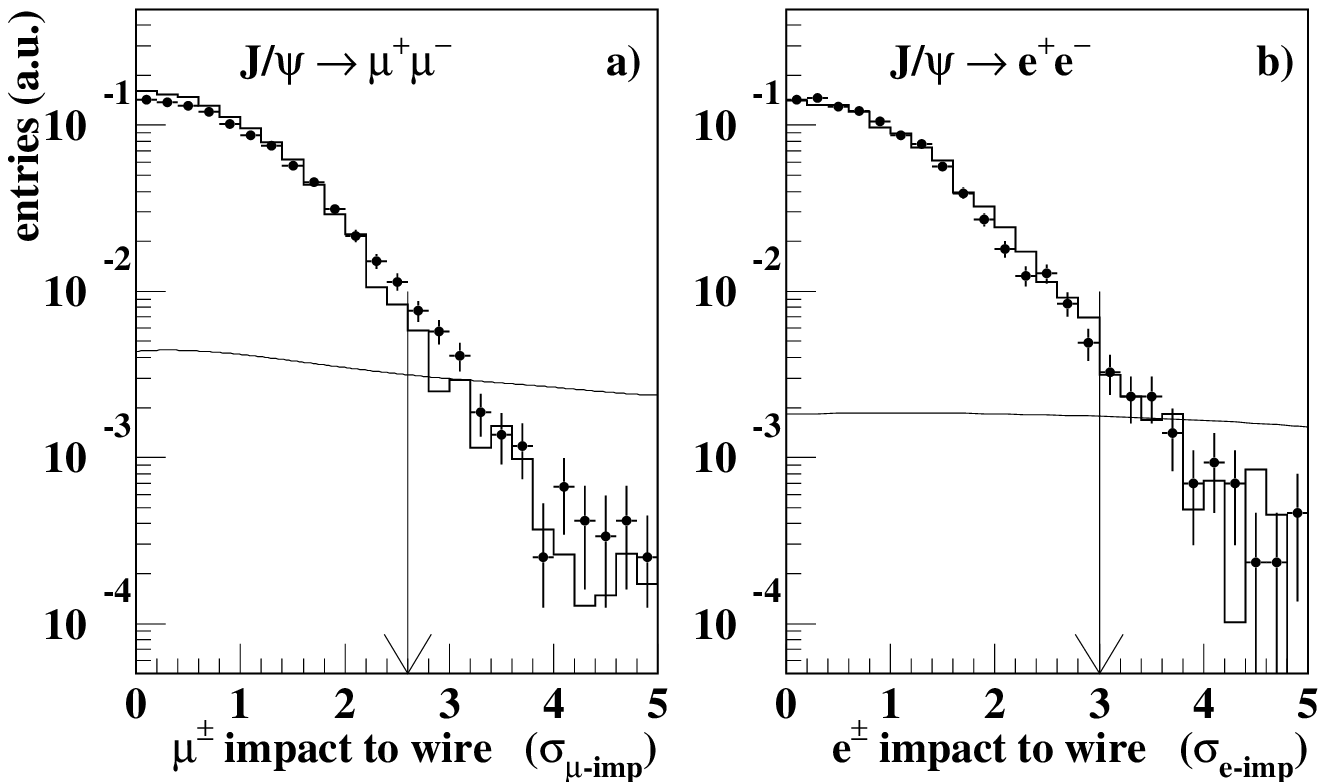}}
\caption{Comparison between the distribution of
the lepton impact parameter to the wire
for real data (points with error bars) 
and for the prompt \jpsi\ MC (histogram). 
a) \jpsimm\ candidates selected in the region
$2.95 < m_{\mm}<3.25$ \mgev\
after sideband subtraction for real data, b)
double BR tagged \jpsiee\ candidates selected in the region
$2.9 < m_{\ee}<3.2$ \mgev\ after sideband subtraction for real data.
The continuous solid line shows the \bjpsi\ MC sample (arbitrary
normalization) while the arrows mark the cut position determined by the
optimization procedure.} 
\myfiglabel{fig:lepimp}
\end{center}
\end{figure}

\subsection{  $\bjpsiX \to \mm X$ }
\label{sect:bjpsimm}

The cut values obtained with the optimization procedure for 
the muon channel are listed in the second
column of Table~\ref{ta:optval}. Once the cuts are applied, 
only 268 events with secondary vertices downstream of the primary interactions
and having a dilepton invariant mass above $2.2~\mgev$ 
survive, while upstream of the target (unphysical region for \bjpsi\ decays),
33 events are found in the same mass region. 
The invariant mass distribution of these detached candidates is shown in
Fig.~\ref{fig:mmdeta}, where a clear peak corresponding to the
\jpsi\ mass is present only in the downstream sample.

\begin{table*}[htbp]
  \begin{center}
    \begin{tabular}{|c|c|c|c|}
      \hline\hline
      Cut                & \mm\ value & \ee\ value & Optimized\\
      \hline
      \hline
      & & & \\[-0.3cm]
      Absolute $\Delta z$  & $> 2~\mbox{mm}$ & $> 2~\mbox{mm}$ & no \\
      \hline
      & & &  \\[-0.3cm]
      $\Delta z$ significance & $> 9.0~\sigma$ & $> 10.0~\sigma$ & yes \\
      \hline
      & & & \\[-0.3cm]
      Lepton impact to wire & $> 2.6~\sigma$ & $> 3.0~\sigma$ & yes\\
      \hline
      & & & \\[-0.3cm]
      \jpsi\ impact to wire   & $ <9.0~\sigma$ & $ <12.0~\sigma$ & yes\\
      \hline
      &  & &\\[-0.3cm]                    
      Lepton identification   & ${\cal L}_\mu >0.05$ & $-2.0~\sigma <E/p-0.99 < 3.5~\sigma$ & $E/p$ only\\
      \hline\hline
    \end{tabular}
  \end{center}
  \caption{Cut values used to select the detached \jpsi\ candidates
           in the muon and in the electron channels. 
           Quantities submitted to the
           blind optimization procedure are explicitly indicated 
           in the last column. }
  \label{ta:optval}
\end{table*}

To count the detached \jpsi\ candidates, an
unbinned maximum likelihood fit is performed using the reconstructed invariant 
mass values.
In the fit, the function used for the prompt \jpsi\
analysis models the peak and an exponential function models the
background. As free fit parameters, we use the \jpsi\ and
background yields and the slope of the exponential.
When the mass position and width of the peak are left as free
parameters, the obtained values are consistent with those of the
prompt \jpsi\ spectrum.
Thus, the final results 
are given with the mass and width fixed to the values
of the prompt signal.
In Fig.~\ref{fig:mmdeta}b, the solid line 
shows the result of the likelihood fit.
For the fit to the total data sample, we find $\nB =46.2^{+8.6} _{-7.9}$ and
$n_\mathrm{bkg} 
= 222 \pm 15$ events for the signal and the background, 
respectively, and a
background slope parameter of $\lambda= -1.9\pm 0.1$ $(\mgev)^{-1}$.

As a cross check, a fit to the upstream events is
performed with the same free parameters. In the full sample, 
the fit finds $\nB=-0.1\pm 1.4$ \jpsi\
events, compatible with absence of signal, $n_\mathrm{bkg}= 33\pm
6$ background events and $\lambda= -1.9\pm 0.3$ $(\mgev)^{-1}$.

\begin{figure}
  \begin{center}
\resizebox{0.48\textwidth}{!}{%
\includegraphics{\Figdir/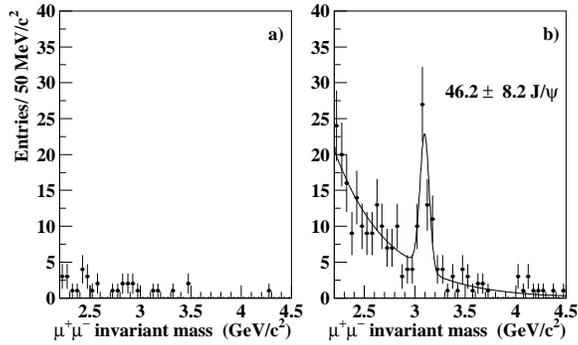}}
\caption{Dimuon mass distributions after vertex detachment cuts, from the full 
muon sample.
a) Upstream events (combinatorial background), b) downstream
events. The solid line shows the result of the likelihood fit.}
\myfiglabel{fig:mmdeta}
\end{center}
\end{figure}

Using the like-sign spectrum and the
background contributions surviving the optimization procedure,
the final composition of the background 
in the detached dilepton spectrum has been estimated. 
Within statistical uncertainty,
all the background can be accounted for by the four background sources 
described above, with the dominant contributions being combinatorial (44$\%$)
and $\bbar$ 
(43$\%$) semileptonic decays.
Since charm particles have a relatively short lifetime, their contribution to
the background is suppressed (13$\%$) by the detachment cuts. 
From MC, no prompt \jpsi\ event ($< 0.7$ at 90\% C.L.)
is expected to survive the detached vertex selection.
Finally, in the region upstream of the target the background is purely 
combinatorial.

To determine $R_{\Delta \sigma}$ in our \xf\ range according to
Eq.~(\ref{eq:dsigbbarss}), the number of prompt \jpsi\ mesons and the efficiency terms
must be evaluated for each of the 14 different sub-samples (one sample for
each wire in each wire configuration, labelled $i$ later). 
The number of prompt \jpsimm\ candidates in each sub-sample is
evaluated with the procedure described in Sect.~\ref{promptmm}. 
To obtain
the efficiencies, the MC events of
prompt \jpsi\ and \bjpsi\ are subjected to the same analysis chain used
for the data. The average final value of the efficiencies is 
$\langle \eff_{R, i} \cdot \eff_{\bbar, i}^{\Delta z}\rangle = 0.400 \pm
0.006$, where the 
uncertainty is given by the statistical fluctuations and sub-sample variability
only.  
All the results obtained
for different target materials are summarized in Table~\ref{tb:xsectratios}. 
The resulting value for $\frac{\dsigbbar}{\Delta\sigma_{\jpsi}}$ in our kinematic range, averaged
over all sub-samples and target materials, is: 

\begin{equation}
R_{\Delta \sigma}(\mm ) = \frac{\dsigbbar}{\Delta \sigma_{\jpsi}} = 
0.0295 \pm 0.0055 \stat
\label{eq:dsigbbmmkin}
\end{equation}
where the uncertainty is statistical only. 
The systematic uncertainties as well as the
tests of the stability of the result are discussed in Sect.~\ref{sec:comb}.

\subsection{  $\bjpsiX \to \ee X$ }
\label{sect:bjpsiee}

As discussed in Sect.~\ref{promptee}, the 
\jpsiee\ decays are affected by a larger background and
the selection of the detached vertex signal is therefore more critical.
Since the electron particle identification (mainly based on the $E/p$
ratio of the lepton tracks) is crucial for candidate
selection, this quantity is included in the blind
optimization procedure.
The cut values obtained for the electron channel are listed in the third
column of 
Table~\ref{ta:optval}.  The Table shows that
the optimized values for the cuts applied in the \ee\ channel
are close to those for the
muon channel, albeit generally stronger.
Only 229 events with dilepton vertices downstream of the primary interactions
and with dilepton invariant masses above $2.0~\mgev$ survive these cuts, 
while 51 events with upstream dilepton vertices are found. 
The invariant mass distribution of these
detached candidates is shown in Fig.~\ref{eedetajpsi:massdetajpsi}; a peak 
is visible in the downstream sample at $m_{\ee}=m_{\jpsi}$.

The same type of unbinned maximum likelihood fit used in the muon channel 
is performed also on the
detached electron candidates, and the fit result is shown in 
Fig. \ref{eedetajpsi:massdetajpsi}b.
For the total data sample, we find $\nB = 36.9^{+8.5} _{-7.8}$ and 
$n_\mathrm{bkg} = 192 \pm 15$ events for the signal and the
background, respectively, and a 
background slope parameter of $\lambda= -1.7\pm 0.1$ $(\mgev)^{-1}$.

As a further cross check,
a fit to the upstream events is performed with the same free parameters.
The fit finds $\nB =-4.2\pm 1.9$
signal events (compatible with absence of signal events), $n_\mathrm{bkg}= 55
\pm
8$ background events and $\lambda= -1.6\pm 0.2$ $(\mgev)^{-1}$.

\begin{figure}
  \begin{center}
\resizebox{0.48\textwidth}{!}{%
\includegraphics{\Figdir/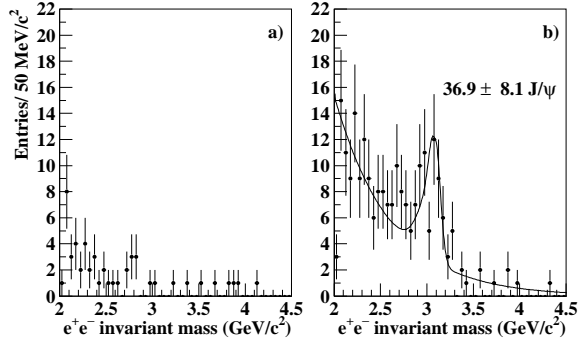}}
\caption{Dilepton mass distribution after vertex detachment cuts,
from the full electron sample.
a) Upstream events (combinatorial background), b) downstream
events. The solid line shows the result of the likelihood fit.
}
\myfiglabel{eedetajpsi:massdetajpsi}
\end{center}
\end{figure}

From the optimization procedure we can also extract the
composition of the remaining background. As in the muon channel,
the four sources considered can account for the obtained spectrum.
The $\bbar$ semileptonic decays
are the dominant contribution (49$\%$), followed by the 
combinatorial background (35$\%$) and the 
open charm events (16$\%$). Also, in this case no 
prompt \jpsi\ event ($< 1.1$ at 90\% C.L.) is expected to
survive the applied cuts for \Bee\  selection.

As for the muon measurement, $R_{\Delta \sigma}$ is evaluated by
determining the number of prompt \jpsi\ mesons and the efficiency
terms separately for each sub-sample (Eq.~\ref{eq:dsigbbarss}). The average
value of the efficiencies is $ \langle \eff_{R, i} \cdot   
 \eff_{\bbar , i}^{\Delta z}\rangle = 0.40 \pm 0.02$, which is close to
 that obtained in the muon analysis.
The resulting value for $\frac{\dsigbbar}{\Delta\sigma_{\jpsi}}$ in the 
electron channel in our kinematic range is:
\begin{equation}
R_{\Delta \sigma}(\ee )  = \frac{\dsigbbar}{\Delta \sigma_{\jpsi}} 
= 0.0353 \pm 0.0078 \stat,
\label{eq:dsigbbeekin}
\end{equation}
in good agreement with the muon result (see Table~\ref{tb:xsectratios}).

\begin{table*}[htbp]
\begin{center} 
\begin{tabular}{|c|ccc|ccc|}
\hline \hline
Channel& \multicolumn{3}{|c@{\ \ \ }|}{\mm} & \multicolumn{3}{|c@{\ \ \ }|}{\ee}\\ 
\hline
Target & Carbon & Titanium & Tungsten & Carbon & Titanium & Tungsten        \\ 
Atomic weight           & 12.01 & 47.87 & 183.84 & 12.01 & 47.87 & 183.84   \\
Prompt \jpsi\ (\nP )    & $93700 \pm 300$ & $8080\pm 100$ &$45560 \pm 200$
                        & $67100 \pm 700$ & $4800 \pm 200$ & $32400 \pm 600$\\
Detached \jpsi\ (\nB )  & $27.8\pm 6.3$ & $3.0\pm 2.1$ & $15.5\pm 4.8 $ 
                        & $17.8^{+5.9}_{-5.2}$ & $0.9\pm 1.0$ & 
                                                       $18.4^{+6.2}_{-5.5}$ \\
$\langle \eff_{R, i} \cdot   \eff_{\bbar , i}^{\Delta z} \rangle$ 
                        &$0.398\pm 0.004$&$0.397\pm 0.011$ &$0.404\pm 0.005$
                        &$0.366\pm 0.007$&$0.424\pm 0.022$ &$0.402\pm 0.014$\\
$\alpha$                & \multicolumn{3}{|c@{\ \ \ }|}{$0.96 \pm 0.01$}
                        & \multicolumn{3}{|c@{\ \ \ }|}{$0.96 \pm 0.01$}    \\
$\Br{\bbjX}$            & \multicolumn{3}{|c@{\ \ \ }|}{$2.32\pm 0.20$}
                        & \multicolumn{3}{|c@{\ \ \ }|}{$2.32\pm 0.20$}     \\ 
\hline
$\dsigB/\dsigP$ ($\times 10^{-2}$) 
                        & $3.22\pm 0.73$ & $4.0 \pm 2.8$  &$3.6 \pm 1.1$  
                        & $ 3.2\pm 1.0 $ & $1.8 \pm 2.2$  &$6.2 \pm 2.0$  \\
$\dsigbbar/\Delta\sigma_{\jpsi}$  ($\times 10^{-2}$)
                        & \multicolumn{3}{|c@{\ \ \ }|}{$2.95\pm 0.55$} 
                        & \multicolumn{3}{|c@{\ \ \ }|}{$3.53\pm 0.78$}     \\ 
\hline \hline
\end{tabular}
\caption{\label{tb:xsectratios} 
Quantities entering into the cross section ratio measurements for the two
channels and for the three target materials. All the numbers are given for the
\xf\ interval $[-0.35, +0.15]$.
}
\end{center} 
\end{table*}


\section{Further  \Bee\ Flavor Confirmation Analyses}
                                                   \label{sec:btag}
To confirm the $b$ flavor content of the selected
detached vertex sample, several independent cross checks have been performed
on  it and on a second sample obtained by relaxing the selection cuts.
The checks include an estimation of the mean lifetime of the detached
candidates. Also a search is made for other decay
products of the \bh s in the enlarged selection, 
with the aim of a complete identification of exclusive decay channels or 
the identification of a potentially cleaner sample of three-prong detached
vertices. 
These and other tests are described in the following.

\subsection{  Lifetime Fit }

The determination of the mean lifetime of the detached candidates
is one of the clearest confirmations of the $b$ flavor of the
signal. A direct and precise measurement is however not possible, 
since in the inclusive \bjpsi\ analysis
the \bh\  is not fully reconstructed and therefore its
momentum and $\beta \gamma$ value are not known. However, 
 MC studies show that a good estimator of
the \bh's $\beta \gamma$ is the corresponding value of the
\jpsi\ meson produced in the decay. Because of this strong correlation,
the proper time $t_i$ of a \Bee\ candidate can be approximated as
$t_i = \Dz_i / (\beta\gamma)_{\jpsi} \cdot k$, where $\Dz_i$ is the decay
length, $(\beta\gamma)_{\jpsi}$ is the $\beta\gamma$ factor
evaluated on the dilepton and $k$ is a numerical factor ($\approx
1.03$) which corrects for the decay and trigger bias. Within the limits of this
approximation, a mean lifetime measurement is therefore possible even without
exclusive identification of \bh s.

\begin{figure}
\begin{center}
\resizebox{0.48\textwidth}{!}{%
\includegraphics{\Figdir/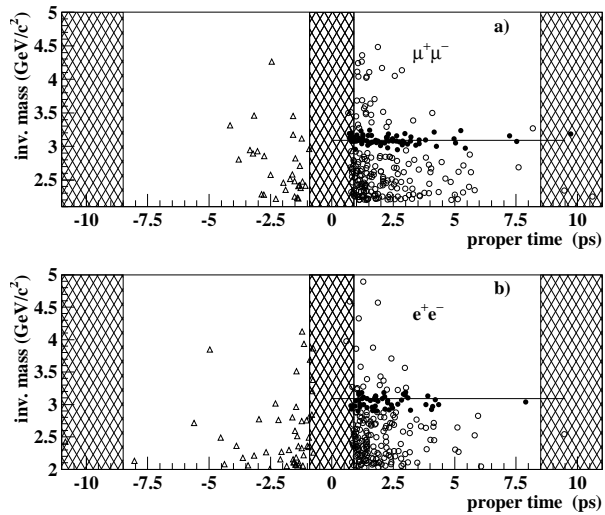}} 
\caption{Scatter plot of \mm\ (a) and \ee\ (b) invariant mass versus
  the proper time for the detached candidates. 
The downstream detached candidates are marked with filled circles in the
\jpsi\ regions ($3.0<m_{\mm}<3.2 \ \mgev$ and
  $2.9<m_{\ee}<3.2 \ \mgev$) and with open circles elsewhere. Upstream
  candidates are shown with open triangle marks.
  The hatches define the regions excluded from the mean lifetime measurement.
}
\myfiglabel{btag:massvsproptime}     
\end{center}
\end{figure}

In Fig.~\ref{btag:massvsproptime}, the
scatter plots of the dilepton invariant mass versus the proper time for the
muon and the electron candidates are shown. As can be seen, 
the proper time distribution behaves differently depending on the invariant mass
region. 
To quantify these differences, unbinned maximum likelihood fits in different
regions have been performed, assuming the usual exponential decay distribution 
for the signal. In these fits, background events are treated like the signal:
only one fit parameter, which represents the mean lifetime of the
decay, is assumed in the likelihood function.
The fit range is limited to the region where trigger and selection
efficiencies are large: 0.9 -- 8.5 ps. Since the efficiency is not constant in
this range, its variation is included in the fitting model, by weighting each
event by the reciprocal of its detached vertex
selection efficiency. 
The results of the fits are summarized in Table \ref{ta:lifetime}.

\begin{table*}
\begin{center}
\begin{tabular}{|lccc|} \hline \hline
Sample & Mass Range (\mgev ) & ~~~$\tau_{\mu\mu}$ (ps)~~~ & ~~~$\tau_{ee}$ (ps)~~~ \\
\hline\hline
data downstream & \jpsi\ region     & $1.61\pm 0.27$ & $1.21 \pm 0.18$ \\
\bjpsi\ MC & \jpsi\ region        & $1.55\pm 0.01$ & $1.57 \pm 0.01$ \\ \hline
data downstream & 2.0--2.5          & $1.30\pm 0.16$ & $0.86 \pm 0.09$\\
data downstream & 3.6--12.0          & $0.55\pm 0.13$ & $0.54 \pm 0.21$\\
data upstream   & 2.0--12.0          & $1.08\pm 0.21$ & $1.16 \pm 0.19$ \\ \hline
\bbar\ semileptonic MC     
& 2.0--2.5          & $1.47\pm 0.12$ & $1.45 \pm 0.04$ \\
\bbar\ semileptonic MC     
& 3.6--12.0          & $0.71\pm 0.09$ & $0.67 \pm 0.03$ \\
$c \bar c$ double semileptonic MC 
& 2.0--12.0          & $0.47\pm 0.06$ & $0.30 \pm 0.02$ \\
\hline\hline
\end{tabular}
\caption{\label{ta:lifetime} 
Results of the unbinned maximum likelihood fits on
  the detached \mm\ and \ee\ candidates in real data 
and on various MC samples. The \jpsi\
  mass range is defined as for Fig.~\ref{btag:massvsproptime}.}
\end{center}
\end{table*}

For detached candidates in the \jpsi\ mass range ($[3.0,3.2]$ GeV/$c^2$
in the muon channel and $[2.9,3.2]$ GeV/$c^2$ in the electron channel),
 the mean lifetime is
1.61$\pm$0.27 ps and  1.21$\pm$0.18 ps for dimuons and dielectrons,
respectively, in agreement with the value obtained from generated and
reconstructed MC \Bee\ decays: 1.56$\pm$0.01 ps.
Background regions in data (side bands in the invariant mass spectrum
and upstream events) usually have lower mean lifetime values and larger
uncertainties due to the poorer statistics.

For reference we also measure the mean lifetime of the main background
channels described by MC. As expected, the open $c\bar c$ events 
generally have a lower
lifetime, while the semileptonic \bbar\ events behave differently
depending on the invariant mass range of the dilepton. The higher mass range is
dominated by combinations of leptons from two different $b$ decay
branches, while, in the low mass range, dileptons from the decay
chain of a single \Bee\ quark dominate. The mean lifetime measured in the
latter sample is clearly close to that of the \Bee .

\begin{figure}[htbp]
\begin{center}
\resizebox{0.48\textwidth}{!}{%
\includegraphics{\Figdir/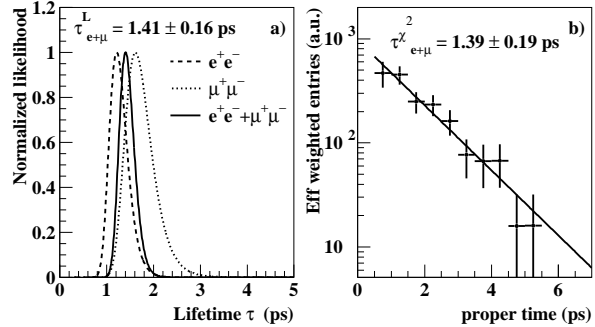}}
\caption{
a) Likelihoods for the \mm and \ee\ detached candidate fits and joint
 likelihood as a function of the mean lifetime;
b) distribution of proper times for the detached \jpsi\ events corrected for
 selection efficiency. 
}
\label{fig:btagjoint}
\end{center}
\end{figure}

To improve the statistical accuracy of the lifetime measurement, a joint
unbinned likelihood fit is performed on the muon and electron candidates (104
events in total). This sample is highly enriched in \Bee\ content, having
about 80 \bjpsi\ candidates plus about 10 events coming from semileptonic
\Bee\ decays.
A value of $\tau^{\cal L} _{e+\mu}=1.41\pm 0.16$~ps is obtained (see
Fig.~\ref{fig:btagjoint}a). Since the total number of candidates is
sufficiently large,
a standard $\chi^2$ fit is also performed on the
efficiency-corrected distribution of proper times for the same events (see
Fig.~\ref{fig:btagjoint}b). In this case, the minimum $\chi^2$ ($\chi^2/\mbox{ndf} =
5.1/8$) is  obtained for $\tau^{\chi^2}_{e+\mu}=1.39\pm 0.19$~ps.
Both results are in good agreement with the expected value for \bh s.

\subsection{ $\jpsi+h^\pm$ analysis }

As described in Sect. \ref{sec:anal}, rather strong (and therefore relatively
inefficient) detachment cuts are necessary to 
do an  inclusive selection of \bjpsi\ 
decays. A partially different sample of $b$ events can be obtained 
by relaxing the detachment cuts and imposing additional requirements on other
quantities which characterize \bh\ decays, for example requiring
the presence of a third track $h^\pm$ forming a good
vertex with the \jpsi\ candidate.
Since the \jpsi\ carries most of the \bh\
momentum, the additional track usually has low momentum and points
far away from the primary interaction. This selection gives access to
$b$ events with smaller detachment from the target wire, but nevertheless
provides high purity because of the stringent requirements on the
three-prong decay vertex (which has an intrinsically lower background).

For events with dimuon candidates selected as described in
Sect. \ref{sec:sel}, 
a search is made for additional decay products among good quality tracks
nearby the \jpsi\ vertex. The presence of a third track, whose closest
distance of approach to the primary vertex is larger than 300 $\mu$m is
required.
Detachment cuts are then applied, and the requirements summarized in Table
\ref{ta:optval} are relaxed. In particular the minimum $\Delta z$ significance 
is reduced to 5$\sigma$, and the minimum lepton impact parameter
to the target wire is lowered to 2$\sigma$.

The dilepton invariant mass and mass vs proper time distribution of
events selected in the muon sample are shown in Fig.~\ref{btag:smafigs}.
A \jpsi\  peak is clearly visible, 
with an improved signal-to-background ratio with respect to the
detached  \jpsi\ analysis of Sect.~\ref{sec:anal}, Fig.~\ref{fig:mmdeta}.
From the unbinned maximum likelihood fit,
the number of  detached \jpsi\ events is $n_{b\bar{b}} = 22 \pm 5$.
Only 50\% of these events are in common with
the previous detached vertex
analysis.

\begin{figure}
  \begin{center}
\resizebox{0.48\textwidth}{!}{%
\includegraphics{\Figdir/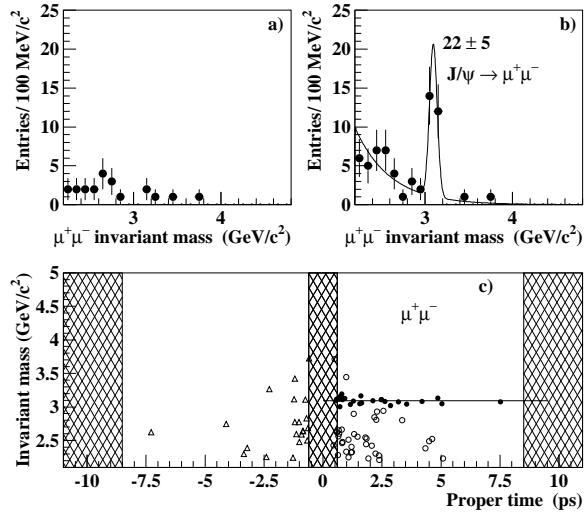}}
\caption{ \mm\ invariant mass for the $\jpsi+h^\pm$ analysis for upstream (a)
  and downstream (b) events. The line in (b) shows
  the result of the unbinned max likelihood fit. c)  \mm\  invariant mass as a
  function of the proper time for the detached candidates.
}
\myfiglabel{btag:smafigs}
\end{center}
\end{figure}

Using these selected events, another value of the cross section
ratio in our acceptance,
$R_{\Delta\sigma} = \dsigbbar/\Delta\sigma_{\jpsi} = 0.043\pm 0.010$,
was obtained.
The measurement is compatible with both the muon and
the electron results presented in Sect. \ref{sec:anal} (within 
0.6$\sigma$ from the electron value, obtained with a 
statistically independent sample).
Since it is partly correlated with the muon result of
Sect.~\ref{sect:bjpsimm}, it is not combined into the final cross section
results presented in Sect.~\ref{sec:comb}.

\subsection{  Other Studies }

Further confirmation of the $b$ flavor content of the detached \jpsi\ 
sample has been obtained from other analyses. 
We studied the kaon population and the distribution of high $p_T$ tracks.
For muon events and for two-BR tag electron events falling in an
invariant mass window around the \jpsi, we observed
that the real data after prompt selection behave as the prompt \jpsi\
MC candidates, while after the detachment selection larger values of both,
the $p_T$ distribution and the kaon population,
are found,
compatible with the \bjpsi\ MC sample.

A search for the exclusive decays \mbox{\bjk} and \bjkp\ has been made.
A few fully reconstructed $B$ meson candidates are found, both for the muon
and the electron channels, but not sufficient for a full exclusive analysis.

The kinematic characteristics of the detached \jpsi\ candidates
have been checked
against the production model employed to evaluate the efficiencies
and were found to be compatible
in our \xf\ and \pt\ acceptance.


\section{Combined Cross Section Measurement}       \label{sec:comb}
The two measurements of $\dsigbbar/\Delta\sigma _{\jpsi}$, presented in
Eqs.~(\ref{eq:dsigbbmmkin}) and ~(\ref{eq:dsigbbeekin}), 
are statistically independent and compatible. 
To have a more precise measurement of
$R_{\Delta \sigma}$, a joint unbinned maximum likelihood fit is performed
on the detached  \mm\ and \ee\  candidates, 
using $R_{\Delta \sigma}$ and all background terms 
($n_\mathrm{bkg}$ and $\lambda$ for the two channels)
as free parameters. 
In Fig.~\ref{res:likeres}, the likelihoods of the invariant mass fits are
shown for \mm\ and \ee\  separately and for the combined fit.
The fit provides the final value
$$R_{\Delta \sigma} = \frac{\dsigbbar}{\Delta \sigma_{\jpsi}} = 0.0314
\pm 0.0049 \stat\ $$ 
where the quoted uncertainty (15\%) is dominated by the statistical
fluctuations on the detached \jpsi\ counting and contains the statistical
contributions from the prompt \jpsi\ 
counting ($\approx 1\%$), the efficiencies ($\approx 5\%$) and  the
$\alpha$ exponent ($\approx 1\%$).

The main sources of systematic 
uncertainties\footnote{
The quoted systematic uncertainties correspond to a $1\sigma$ interval and are
  usually evaluated as the maximal interval width of the cross section 
variation divided by $\sqrt{12}$.}, which are not related to the
\bbar\ statistics, are due to the $\Br{\bbjX}$ (8.6\%), to the trigger and
reconstruction efficiency ratio \effR\ (5\%), to the \Bee\ production and
decay model (5\%), to the prompt \jpsi\ production model (3.1\%) and to the
prompt \jpsi\ counting (1.5\%). Other contributions are below the 1\% level.
To determine the sensitivity of the results to the
cut values, we vary them within reasonable ranges, always requiring a
negligible prompt \jpsi\ background. 
A conservative estimate of 5\% systematic uncertainty has been obtained.
The uncertainties in the background shape and yield give a negligible
contribution to the systematic uncertainty (below 1\%) in the muon channel but
a sizeable contribution in the electron channel
(7\%). The overall systematic uncertainty for the two channel average
measurement is 14\%.
\begin{figure}
 \begin{center}
\resizebox{0.48\textwidth}{!}{%
\includegraphics{\Figdir/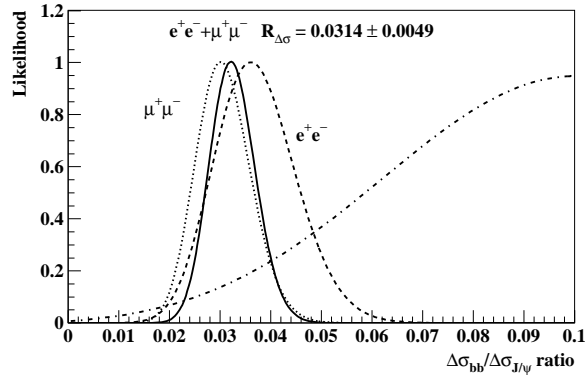}} 
\caption{Likelihoods of the invariant mass fits as a function of
$R_{\Delta \sigma}$, shown separately for \mm\ data (dotted line), \ee\ data
(dashed line) and for the combination of the two (continuous line). 
The dot-dashed line shows the likelihood function corresponding to the 2000
data analysis ~\cite{HBbb} which is used in the final average.
}
\myfiglabel{res:likeres}
\end{center}
\end{figure}

The weighted averages of the muon and electron 
results for the different target materials and for the total are presented in
Table~\ref{tab:bbvsA}.   As can be seen from the $\chi^2$ probability, 
the two channels are always compatible. The three $\frac{\dsigB}{\dsigP}$
measurements are also compatible with the assumed linear $A$ dependence of
the \bbar\ cross section.

The present result is $1.7 \sigma \stat$ below the previous \hb
measurement~\cite{HBbb} 
($R_{\Delta\sigma}=0.107 \pm 0.047 \stat \pm 0.022 \sys$) 
which was obtained with a similar technique but with
a much smaller data sample (dot-dashed line in Fig.~\ref{res:likeres}).
Since the two data samples are independent, we
combine the two measurements assuming
that the systematic errors are uncorrelated.
We obtain:

\begin{equation}
R_{\Delta \sigma} = \frac{\dsigbbar}{\Delta \sigma_{\jpsi}} = 0.032
\pm 0.005 \stat \pm 0.004 \sys .
\label{eq:resrdsfin}
\end{equation}

\begin{table*}[htb]
\begin{center} 
\begin{tabular}{|l|c|c|cc|c|}
\hline \hline
Material (A) &$N_{\jpsi}$ & $\nB$ & $\sigma$ ratio & value &Prob($\chi^2$) \\
\hline \hline
Carbon (12)& 161 k & $45.6 \pm 8.4$  & 
&$0.0324\pm 0.0059$ & 94\% \\
Titanium (48) & 13 k  & $ 3.9 \pm 2.3$  & 
$\dsigB/\dsigP$ & $0.027\pm 0.018$     & 54\% \\
Tungsten (184)& 78 k  & $33.9 \pm 7.6$  & 
&$0.044 \pm 0.010$    & 29\% \\ \hline
Total      & 251k  & $83   \pm 12$  & 
$\dsigbbar/\Delta\sigma_{\jpsi}$ & $0.0314\pm 0.0049$   & 54\% \\
\hline \hline
\end{tabular}
\caption{\label{tab:bbvsA} 
Combined \mm\ and \ee\  results: sum of the number of prompt ($N_{\jpsi}$) 
and detached \jpsi\ ($\nB$) found for the two channels together,
 $\dsigB/\dsigP$ (rows 1--3 ), $\dsigbbar/\Delta\sigma_{\jpsi}$ (row 4),
and $\chi^2$ probability of the average value.
}
\end{center} 
\end{table*}

To compare our result with other measurements and theoretical
predictions, we extrapolate the $R_{\Delta \sigma}$ ratio to
the full kinematic range and then make use of the value of $\sigma_{\jpsi}$
presented in Sect.~\ref{sec:meth}, obtaining the \sigbbar\ value:
\begin{equation}
\sigbbar = 14.9 \pm 2.2 \stat \pm 2.4 \sys \ \textrm{nb/nucleon}.
\label{eq:ressigma}
\end{equation}

This result can be compared with the available measurements, which
were obtained in $p$Au \cite{E789bb} and $p$Si \cite{E771bb} collisions at 800
\pgev\ proton momentum. Due to the near-threshold conditions in which all
experiments (including \hbp) operated, it is necessary 
to rescale the Fermilab results for the different $\sqrt{s}$ value of the
collision. The latest QCD calculations predict an increase of the \bbar\ cross
section between 800 (Fermilab) and 920 \pgev\ (\hbp ) by
$(42 \pm 9)\%$
\cite{nason1998,vogt2004}. All the available measurements are presented in
Table~\ref{tab:finbb}, where a scaling of the Fermilab measurements 
to \hb energies
is
performed.
As can be seen, the present result (Eq.~\ref{eq:ressigma}) is
consistent with both the E789 value (within 1.6$\sigma$) and with the E771
value (within 1.8$\sigma$).

\begin{table*}[htb]
\begin{center}
\begin{tabular}{|l|c|c|c|c|c|c|c|}
\hline \hline
Expe-   &      &       & $p$   &  & $\sigma(b \bar b)$  & 
$\sigma(b \bar b)$ & \\ 
riment & Year & Target & (\pgev) & Events & (nb/nucleon) &
at 920 \pgev & Ref. \\ \hline \hline
E789 & 1995 & Au & 800 & 19 &$5.7 \pm 1.5 \pm 1.3$ &
$8.1\pm 2.2\pm 1.9$  & \cite{E789bb}\\
E771 & 1999 & Si & 800 & 15 & $43^{+27}_{-17} \pm 7$ & 
$61 ^{+38} _{-24} \pm 11$ & \cite{E771bb}\\ 
\hb & 2002 & C/Ti & 920 & 10 & $32^{+14\ +6}_{-12\ -7} \ ^{(*)}$ & 
$32^{+14\ +6}_{-12\ -7} \ ^{(*)}$ & \cite{HBbb}\\ 
\hb & 2005 & C/Ti/W & 920 & 83 & $14.4 \pm 2.2 \pm 2.3$ & 
$14.4 \pm 2.2 \pm 2.3$ & t.w.\\ \hline \hline
\hb & 2002/5 & C/Ti/W & 920 & 93 &$14.9 \pm 2.2 \pm 2.4 $ & 
$14.9 \pm 2.2 \pm 2.4 $ & t.w.\\ \hline \hline

\end{tabular}
\caption{\label{tab:finbb} 
Present experimental situation on the 
\bbar\ cross section in pN interactions with the comparison of all the
measurements rescaled at 920 \pgev\ proton momentum.\\
$^{(*)}$ Value as quoted in the paper, which was obtained using a
$\sigma_{\jpsi}$ value different from the current paper.}
\end{center} 
\end{table*}

The measurements presented in Table~\ref{tab:finbb} can also be compared to
the latest QCD predictions~\cite{nason1998,vogt2004},
which have been performed by two theory
groups. They both employ Next-to-Leading-Order (NLO) calculations
with resummation techniques to take into account the large corrections due to
emission of soft gluons. 
The calculations give $\sigbbar\ = 28 \pm 15$ nb/nucleon \cite{vogt2004} and
$\sigbbar\ = 25^{+20}_{-13}$ nb/nucleon \cite{nason_priv}: the results are
shown in Fig.~\ref{fig:bbarexpthend} as a function of \mbox{$\sqrt{s}$. }
As can be seen, the present measurement is compatible with the theoretical
calculations. 

\begin{figure}
\begin{center}
\resizebox{0.48\textwidth}{!}{%
\includegraphics{\Figdir/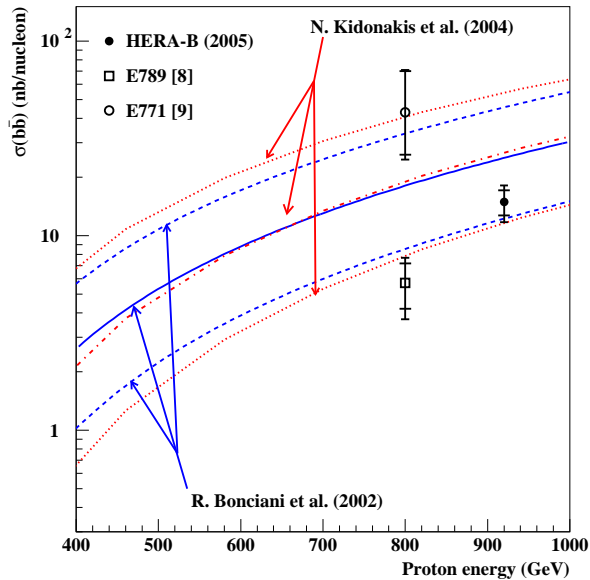}}
\caption{Comparison of the available \sigbbar\ measurements  
with the theoretical predictions of R. Bonciani {\it et al.} \cite{nason1998} 
updated with the NNLO parton distribution function in \cite{mrst}
(solid line: central value, dashed lines: upper and lower bounds) and
N. Kidonakis {\it et al.} \cite{vogt2004} (dot-dashed line: central value, 
dotted lines: upper and lower bounds). The upper and lower bounds have been
defined by changing the renormalization and factorization scales (by a factor
0.5 and 2.0) and the \Bee\ quark mass ($4.5-5.0$ \mgev ).
}
\myfiglabel{fig:bbarexpthend}    
\end{center}
\end{figure}


\section{Conclusions}                              \label{sec:concl}
A search for $\bjpsiX \to l^+ l^- X$ decays has been performed on a sample of
164 million dilepton trigger events, acquired during the \hb 2002-2003
physics run. The data analysis, based on identification of detached 
vertices, resulted in $46.2\pm 8.2$ \ $\bjpsiX \to \mm X$ 
candidates and $36.9 \pm 8.1$ \ $\bjpsiX \to \ee X$ candidates.

From these events, we measure the ratio $\dsigbbar/\Delta\sigma_{\jpsi}$ 
of the \bbar\ production cross section to the prompt \jpsi\ cross section in
the \hb\ acceptance range ($ -0.35 < \xf < 0.15 $).
Within statistical uncertainties, the results for the muon and for the electron
channel  are compatible and yield a combined value of
 $\dsigbbar/\Delta\sigma_{\jpsi} = 0.0314\pm 0.0049 \stat \pm 0.0044 \sys$, 
 where the systematic  uncertainty (14\%) is dominated by the  8.6\% 
 contribution 
 due to the branching fraction \Br{\bbjX}. The present measurement is then 
 combined with the \hb 2000 measurement \cite{HBbb}, 
for a final result of:

\begin{equation}
R_{\Delta \sigma} = \frac{\dsigbbar}{\Delta \sigma_{\jpsi}} = 0.032
\pm 0.005 \stat \pm 0.004 \sys.
\end{equation}
 
Further investigations of the detached vertex sample confirm the \Bee\ 
nature of the candidates.
The measured mean lifetime $\tau=1.41\pm 0.16$ ps agrees well
with previous measurements of the average $\Bee$ lifetime~\cite{PDG}. 
A search for extra
tracks coming from the detached \jpsi\ vertex in the muon channel 
yields $22 \pm 5$ partially reconstructed $\Bee \to \jpsi+h^\pm +X$
events. From this latter sample, 50\% of which consists of detached \jpsi\
mesons not found in the inclusive analysis, a $R_{\Delta
  \sigma}$ cross section ratio in good agreement with the main
measurement is obtained. 
In addition to these checks, we have verified that the detached candidate
events have a kaon and \pt\ distribution that is in good agreement with that
expected for $\Bee$ events.

To compare our result to previous measurements and to theoretical calculations,
we compute the \sigbbar\ cross section by
extrapolating the ratio $R_{\Delta \sigma}$ to the full phase space and
using the value of $\sigma_{\jpsi}= (502 \pm 44)\
\mathrm{nb/nucleon}$
\cite{refjpsi}.
The resulting \sigbbar\ cross section is 
$14.9 \pm 2.2 \stat \pm 2.4 \sys \ \textrm{nb/nucleon}$.
Comparing to the other available experimental results, the present value is 
within 1.6$\sigma$ of the E789 value \cite{E789bb} (after rescaling
to the same $\sqrt{s}$) and 1.8$\sigma$ below the rescaled E771 measurement
\cite{E771bb}. 

The \hb\ value can be compared with the latest QCD 
predictions, calculated beyond the NLO order and considering 
the effects of soft-gluon re-summation, performed at 920 \pgev\ proton momentum.
Our result is within 1$\sigma$ of the current calculations \cite{vogt2004,nason_priv}.


\begin{acknowledgments}
\setcounter{footnote}{2}
\renewcommand\thefootnote{\fnsymbol{footnote}}
We express our gratitude to the DESY laboratory for the strong support in
setting up and running the \hb  experiment. We are also indebted to the
DESY accelerator group for the continuous efforts to provide good and
stable beam conditions. 
The \hb experiment would not have been possible without the enormous
effort and commitment of our technical and administrative staff. It is a
pleasure to thank all of them. 

In the preparation of this paper we have benefitted from many useful
discussions with M.~Mangano, N.~Kidonakis, P.~Nason and R.~Vogt 
on the theory of \bbar\ production, and with F.~Maltoni on the \jpsi\
hadroproduction models. 
\end{acknowledgments}

%

\end{document}